\documentstyle[12pt]{article}

\newcommand{\be}{\begin{equation}}
\newcommand{\ee}{\end{equation}}
\newcommand{\ba}{\begin{eqnarray}}
\newcommand{\ea}{\end{eqnarray}}

\newcommand{\eq}[1]{(\ref{#1})}
\newcommand{\Eq}{Eq.~\eq}
\newcommand{\Sect}[1]{Sect.~\ref{#1}}

\newcommand{\ra}{\rangle}
\newcommand{\la}{\langle}

\begin{document}

\title{Quantum dissipative systems from theory of continuous measurements}
\author{{Michael B. Mensky}\\[2mm]
P.N.Lebedev Physics Institute, 117924 Moscow, Russia\\[1mm]
\and {Stig Stenholm}\\[2mm]
Royal Institute of Technology (KTH), Stockholm}

\date{}

\maketitle

\begin{abstract}
We apply the restricted-path-integral (RPI) theory of non-minimally
disturbing continuous measurements for correct description of
frictional Brownian motion. The resulting master equation is
automatically of the Lindblad form, so that the difficulties typical
of other approaches do not exist. In the special case of harmonic
oscillator the known familiar master equation describing its 
frictionally driven Brownian motion is obtained. A thermal reservoir 
as a measuring environment is considered. 
\end{abstract}

\section{Introduction}\label{SectIntro}

Irreversible behavior is an every day phenomenon in physics, even if the
basic theories have simple properties under time reversal. It has long been
realized, that in certain limits, an incoherent environment will provide a
reservoir for relaxation processes, which in thermal equilibrium will
enforce the proper distribution functions. More recently, it has been
realized that also observations by measurements introduce decoherence, which
from the point of view of the system evolution is equivalent to
environmental dephasing. A comprehensive summary of the
recent progress in this field is found in Ref.~\cite{bkZeh}.

In the case of a continuous but weak observation, we can design quantum
probability amplitudes conditioned on the continuously observed variable. 
Such a system can be formulated in a natural way by putting
restrictions on the path integral providing the quantum propagator of the
system. Such restricted path integrals (RPI) have been found to clarify many
questions related to the continuous monitoring of a selected variable
\cite{MBMbk00}. In particular, it can incorporate the unavoidable
disturbing effect of the observation back on the system observed.
This is termed a minimally disturbing measurement.
Non-minimally disturbing continuous measurements have also been 
suggested \cite{MBMbk00}.
In this paper we pursue the consequences of a simple model for such
a measurement.

The RPI approach is basically selective. This means that the
measurement readout, or history of observations, is taken into account
so that the dynamics of the measured system conditioned by this
history is presented by a state vector.
It is possible to connect the path integral formulation to the more common
master equation approach \cite{MBMbk00}.
This is done by going over to the non-selective description of
the continuous measurement. If we take the time evolution conditioned
by the given measurement readout (history of observations), express it
in terms of a density matrix and then
perform a functional summation over all possible observed
histories, we obtain an ensemble description comprising all possible
histories of observations. The resulting total density matrix
obeys a master equation.

The approach to environmental influences making use of master
equations goes back a long time, but the recent interest has been
kindled by the works of Caldeira and Leggett, Zurek, Unruh and
Zurek \cite{MasterEq}. In many connections, especially those related
to quantum optics, these results are satisfactory, but when applied to
frictional Brownian motion they are found to be flawed
\cite{r6}; see also the discussions in \cite{DifficultFrict}. The
reason is that they are not of the Lindblad form, and hence they
cannot preserve the positivity of the density matrix. It is important
that this is the case even if the fluctuation-dissipation theorem is
fulfilled. Some authors claim that this is of little
consequence, but it does signal a possible danger in
applications. The point is that even correctly derived master
equations fulfilling the fluctuation-dissipation theorem, may not be
of the Lindblad form \cite{Lindblad76}. A discussion of the situation
is found in Ref.~\cite{Stenholm96}.

Here we apply the RPI theory of continuous measurements to
obtain correct description of frictional Brownian motion. We
formulate the RPI description for a non-minimally disturbing
continuous observation. Obtaining the phenomenological description of
the measurement (including its back-action on the measured system), we
go over to the non-selective description and investigate the resulting
master equation. It turns out to be of Lindblad form for any choice of
the parameters of the measurement.

Then we derive the equations of motion for the first and second
moments which turn out to nearly coincide with the expected
Ornstein-Uhlenbeck equations. The main correction is a position
diffusion coefficient which may be derived from the quantum
uncertainty relation. The conditions are found when this term may be
omitted. In case of thermal equilibrium of the measured
system with thermal reservoir, the strength of the continuous
measurement performed by the reservoir is shown to depend on the
reservoir's temperature. If, however, the interaction between the
system and reservoir is weak enough, the equilibrium turns out to be
impossible, and the relation between the strength of the measurement
and the temperature is violated.

The paper is organized in the following way. In \Sect{SectNon-Min} we
formulate the RPI description of a (non-minimally disturbing)
continuous measurement, derive the corresponding master equation and
show that it automatically turns out to be of the Lindblad type. In
\Sect{SectHarmOscil} we restrict our treatment to the familiar case of
a harmonic oscillator. We find that continuously monitoring the
momentum of the oscillator (by its environment) leads to the familiar
form of the master equation \cite{Stenholm96} describing the
frictionally driven Brownian motion. This gives the
Ornstein-Uhlenbeck-type equations for the first and second moments. In
\Sect{SectReservoir} the equilibrium of the measured system with the
(measuring) reservoir having a definite temperature is discussed.
Finally \Sect{SectConclud} summarizes and concludes our paper.

\section{Non-minimally disturbing measurement}\label{SectNon-Min}

The restricted path integral (RPI) for the {\em monitoring of a
quantum observable $A(q,p)$} with the observed history $a(t)$ may be
formally written \cite{MBMbk00} as an ordinary Feynman path integral
(in the phase-space representation) with the weight functional of the
form
\begin{equation}
w_0[a]=\exp \left\{ \int dt
\left[ -\kappa \left( A(q,p)-a(t)\right) ^2\right]
\right\}   \label{a1}
\end{equation}
included in the itegrand. In case of {\em non-minimally disturbing
monitoring}, imaginary terms are also present in the exponent
\cite{MBMbk00}. In the {\em linear approximation} this gives
\begin{equation}
w[a]=\exp \left\{ \int dt
\left[ -\kappa \left( A(q,p)-a(t)\right) ^2
-\frac i\hbar \left( \lambda \,a(t)B(q,p)+C(q,p)\right)
\right]
\right\}
\label{NonMinWeight}\end{equation}
Here $\kappa$ and $\lambda$ are the real parameters which characterize
correspondingly the strength of the measurement (monitoring) and
non-minimal disturbance produced by it. The real term in the exponent
leads to the restriction of the path integral
that is analogous to von Neumann's projecting in case of instantaneous
measurements and is necessary for observations. Imaginary terms are
responsible for adding phase in the course of the measurement which is
not necessary for the observation, hence the terminology
`non-minimally disturbing measurement'.

With these definitions, the time evolution conditioned on the
observation of $a(t)$ can be written as a RPI
\ba
U_{[a]}(q,q',t) & = &
\int d[p] \int_{q}^{q'} d[q]\exp \left\{ \int_0^tdt\left[
\frac i\hbar \left( p\,\dot{q}-H(q,p)\right) \right. \right. \nonumber \\
&  & \left. \left. -\kappa \left( A(q,p)-a(t)\right) ^2-\frac i\hbar \left(
\lambda \,a(t)B(q,p)+C(q,p)\right) \right] \right\} .
\label{a3} 
\ea
The corresponding evolution operator $U_{[a]}(t)$ allows one to
express the density matrix at an arbitrary time
conditioned on the observation of $a(t)$:
\be
\hat{\rho}_{[a]}(t)=U_{[a]}(t)\,\hat{\rho}(0)\, U_{[a]}^{\dagger}(t) .
\label{CondEvol}\ee

The ensemble averaged or {\em total density matrix}
is obtained from this by carrying out the functional
integration over all possible  observations (measurement readouts) $[a]$:
\be
\hat{\rho}(t)=\int d[a]\, \hat{\rho}_{[a]}(t)
\label{TotEvol}\ee

As a result we obtain
\ba
\hat{\rho}(t) 
& = & \int d[q',p']\int d[q'',p'']\;\rho _0(q',q'') \nonumber\\
&\times& \exp 
\left[ \int_0^t dt\left( \frac i\hbar \left( p'\dot{q}'-H'-C'
\right) -\frac i\hbar \left( p''\dot{q}''-H''-C''\right) 
\right.\right. \nonumber\\
&-&\frac \kappa 2\left( A'-A''\right)^2
-\frac{\lambda^2}{8\kappa\hbar ^2}\left( B'-B''\right)^2
\left. +\frac{i\lambda }{2\hbar }\left( A'+A''\right) 
\left( B''-B'\right) \right] 
\label{a5}
\ea
where the notations
$A'=A(q',p')$, $B''=B(q'',p'')$ etc. are used.

From this we can calculate the master equation for the density
matrix by taking the time derivative and noting that all primed
operators go to the left of $\hat{\rho}$ and the doubly primed ones to
the right.

We still need to {resolve the order of} $\hat{A}$ and $\hat{B}$ when
taken at the same time. Our physical interpretation suggests a
solution: as $\hat{B}$ is the back action of observing $\hat{A}$ , the
latter should act before the former. We thus obtain the equation
\ba
\frac \partial {\partial t}\hat{\rho} 
& = & -\frac i\hbar \left[ \hat{H}+\hat{C},\hat{\rho}\right] 
-\frac \kappa 2\left[ \hat{A},\left[ \hat{A},\hat{\rho}\right] \right] 
-\frac{\lambda ^2}{8\kappa \hbar ^2}
\left[ \hat{B},\left[ \hat{B},\hat{\rho}\right] \right] \nonumber \\
&  & -\frac{i\lambda }{2\hbar }
\left[ \hat{B},\left[ \hat{A},\hat{\rho}\right] _{+}\right] .
\label{a6}
\ea

Let us rewrite the master equation \eq{a6} by introducing the operator
\begin{equation}
\hat{l} =
\hat{A}-i\frac{\lambda}{2\kappa\hbar}\hat{B}  
\label{a7}
\end{equation}
Solving for $\hat{A}$ and $\hat{B}$ and inserting into \Eq{a6} we obtain
\ba
\frac \partial {\partial t}\hat{\rho}
& = & -\frac i\hbar 
\left[ \hat{H}+\hat{C}
-i\frac{\kappa\hbar}{4}\left( \hat{l}^{\dagger 2}-\hat{l}^2\right)\, , 
\, \hat{\rho}\, \right]  \nonumber \\
&  &  \nonumber \\
&  & - \frac{\kappa}{2} \left(
\hat{l}^{\dagger }\hat{l}\,\hat{\rho}
-2\,\hat{l}\,\hat{\rho}\,\hat{l}^{\dagger }
+\hat{\rho}\,\hat{l}\,\hat{l}^{\dagger }\right)
\label{a8}
\ea
The resulting equation is of the Lindblad form. The original
Hamiltonian of the measured system is renormalized by the measurement
procedure.

\section{Special case of a harmonic oscillator} \label{SectHarmOscil}

We shall now specialize to the case of a harmonic oscillator,
\begin{equation}
\hat{H}=\frac{\hat{P}^2}2+\frac 12\omega ^2\hat{Q}^2.  
\label{a10}
\end{equation}
Let the {\em momentum} operator be monitored
and the {\em coordinate} operator presents non-minimal disturbance
(this means that the momentum is shifted when being monitored):
\begin{equation}
\hat{A}=\hat{P}, \quad \hat{B}=\omega \hat{Q}  \label{a9}
\end{equation}
The weight factor \eq{NonMinWeight} takes the form
\begin{equation}
w[a]=\exp \left\{ \int dt
\left[ -\kappa \left( p-a(t)\right) ^2
-\frac i\hbar \left( \lambda\omega \, a(t) \, q+C(q,p)\right)
\right] \right\} .
\label{NonMinWeightOsc}\end{equation}
Exponential of operator $\hat Q$ is the displacement operator for
observable $\hat P$. Therefore, the non-minimal disturbance (determined
by the term proportional to $q$ in \Eq{NonMinWeightOsc}) consists in
this model in shifting the momentum (the same observable which is
measured).

The master equation now becomes (for $\hat{C}=0$)
\ba
\frac \partial {\partial t}\hat{\rho} 
& = & -\frac i\hbar \left[ \hat{H},\hat{\rho}\right] 
-\frac \kappa 2\left[ \hat{P},\left[ \hat{P},\hat{\rho}\right] \right] 
-\frac{\lambda ^2\omega ^2}{8\kappa \hbar ^2}
\left[ \hat{Q},\left[ \hat{Q},\hat{\rho}\right] \right] \nonumber \\
&  & -\frac{i\lambda \omega }{2\hbar }
\left[ \hat{Q},\left[ \hat{P},\hat{\rho}\right] _{+}\right] .
\label{a11}
\ea
This is the well known master equation for a {\em Brownian motion}
\cite{Stenholm96}.
The Brownian motion of the oscillator is thus interpreted as
the {\em effect of monitoring the momentum}
by a continuously acting environment (reservoir).

The equations for the first moments (mean values of $\hat{P}$ and
$\hat{Q}$) following from \Eq{a11} are
\ba
\langle \frac{\partial \hat{P}}{\partial t}\rangle  
& = & -\omega ^2\langle\hat{Q}\rangle 
-\lambda \omega \langle \hat{P}\rangle \nonumber  \\
\langle \frac{\partial \hat{Q}}{\partial t}\rangle  
& = & \langle \hat{P}\rangle
\label{a12}
\ea
and the equations for the {\em second moments}
\ba
\frac \partial {\partial t}\langle \hat{P}^2\rangle  
& = & -\omega^2\langle \hat{P}\hat{Q}+\hat{Q}\hat{P}\rangle 
-2\lambda \omega \langle \hat{P}^2\rangle 
+\frac{\lambda ^2\omega ^2}{4\kappa } \nonumber \\
\frac \partial {\partial t}\langle \hat{P}\hat{Q}+\hat{Q}\hat{P}\rangle  
& =& -\lambda \omega \langle \hat{P}\hat{Q}+\hat{Q}\hat{P}\rangle 
+2\left(\langle \hat{P}^2\rangle -\omega ^2\langle \hat{Q}^2\rangle \right) 
\nonumber\\
\frac \partial {\partial t}\langle \hat{Q}^2\rangle  
& = & \langle \hat{P}\hat{Q}+\hat{Q}\hat{P}\rangle +\kappa \hbar ^2.
\label{a13}
\ea
We can introduce a damping constant in Eqs.(\ref{a12}) and (\ref{a13}) by
writing $\gamma =\lambda \omega$. Equations \eq{a13} are appropriate for an Ornstein-Uhlenbeck process
except for the {\em diffusion term} in position space. The same type
of equations follow from the model introduced by Gallis \cite{r10};
see also \cite {Stenholm96}. Also Di\'{o}si \cite{Diosi} has
pointed out that such a term is necessary to obtain a Lindblad form of
evolution.

In the classical limit, $\hbar \rightarrow 0$,  Eq. (\ref{a13}) gives 
{\em the steady-state solution}  
\be
\langle \hat{P}\hat{Q}+\hat{Q}\hat{P}\rangle =0, \quad
{\langle \hat{P}^2\rangle }=\omega ^2\langle \hat{Q}^2\rangle 
=\frac{\lambda\omega}{8\kappa}
\label{SteadyState}\ee
which agrees with the virial theorem. These result in the expression for the
mean energy 
\be
\la{\hat H}\ra=\frac 12 \la \hat P^2 + \omega^2\hat Q^2 \ra
= \frac{\lambda\omega}{8\kappa}.
\label{MeanHamilt}\ee

The additional diffusion term in the third of Eqs.~(\ref{a13}) has got no
simple physical interpretation within a classical stochastic model. However,
it is easily {\em interpreted in the scheme of measurements} as a
consequence of the uncertainty relations. The fact that the term is
proportional to $\hbar ^2$ signals its quantum origin. Indeed, with time
passing, momentum is measured with better precision, and therefore the
uncertainty of the coordinate becomes larger. This may be shown to be
expressed by just this term; see \cite{MBMNamiot} and \cite[Sect.~4.3.4]
{MBMbk00} for the dual situation when the coordinate is continuously
measured and grows more uncertain.

This diffusion term causes a spreading of the position variable according to 
$\Delta \langle \hat{Q}^2\rangle \approx \kappa \hbar ^2t.$ If this is
required to be much smaller than the steady state value in 
(\ref{SteadyState}), we find that its effect is negligible if 
\begin{equation}
t\ll \frac \lambda {8\kappa ^2\omega \hbar ^2}.  \label{DiffusNeglig}
\end{equation}

\section{Thermal reservoir as an environment} \label{SectReservoir}

Let us assume that the environment of our (measured) oscillator is a
thermal bath at some temperature $T$ and the oscillator is in
equilibrium with the environment. Then the density matrix of the
oscillator corresponds to the Boltzmann distribution in energies,
which may be used for calculating mean value of the number of energy
level $\hat n=a^{\dag} a$ and therefore mean value of energy:
\be
\bar{n}=\frac 1{\exp \left( \hbar \omega /k_BT\right) -1},
\quad
\la{\hat H}\ra=\hbar\omega\la \hat n + \frac 12\ra
=\frac{\hbar\omega}{2}\coth\frac{\hbar\omega}{2k_BT}
\label{PartNumber}\ee
This provides the {\em fluctuation-dissipation theorem} which relates 
the momentum diffusion coefficient and friction coefficient in the form
\begin{equation}
D=\frac{\gamma \hbar \omega }2
\coth \left( \frac{\hbar \omega }{2k_BT}\right) .  \label{a14}
\end{equation}
This gives in our case 
\begin{equation}
\lambda =4\kappa \hbar
\coth \left( \frac{\hbar \omega }{2k_BT}\right).
\label{lambda2}\end{equation}
Therefore, the strength of non-minimal disturbance depends on
temperature of the measuring reservoir.

From \Eq{lambda2} the inequality
\be
\lambda > 4\hbar\,\kappa 
\label{lambda3}\ee
follows. We see from this inequality that for $\lambda>4\hbar\,\kappa$, 
the oscillator may be in {\em thermal equilibrium} with the reservoir.
It can be shown that the {\em regime of measurement is in this case
classical}, quantum effects are negligible.

The question naturally arises as to what happens if the non-minimal
disturbance of the measurement is small enough (in comparison with the
strength of the measurement) so that $\lambda<4\hbar\,\kappa$.
In this case the measurement is performed in {\em quantum regime}, but
the situation seems to contradict the inequality \eq{lambda3}. However,
we have derived this inequality under the assumption that the
distribution of the oscillator over energies is of the Boltzmann
form, i.e. the oscillator is in equilibrium with the thermal
reservoir. If this inequality is violated, thermal equilibrium of the
oscillator with the thermal reservoir is impossible.

All these conclusions are made in the framework of the certain model
of the continuous measurement. In this model the momentum is
continuously monitored and non-minimally disturbed (see the remark
following \Eq{NonMinWeight}). Because of symmetry between momentum and
coordinate, the same is valid also if the coordinate is measured and
non-minimally disturbed. The conclusions are however not justified if
the oscillator is continuously measured but according to another
model.

It is also interesting to evaluate the energy shift caused by the
observational procedure. Reading off this shift from Eq.(\ref{a8}) and
making use of \Eq{a9} we obtain
\begin{equation}
-i\frac{\kappa\hbar}{4}\left( \hat{l}^{\dagger 2}-\hat{l}^2\right)
=\frac{\lambda \omega }4\left( \hat{P}\hat{Q}+\hat{Q}\hat{P}\right) .
\label{a18}
\end{equation}
Combining this with the Hamiltonian (\ref{a10}) we obtain
\ba 
\hat{H}_{\rm eff}
&=& \frac 12\left(\hat{P}^2+\omega ^2\hat{Q}^2\right)
+\frac{\lambda \omega }4\left( \hat{P}\hat{Q}+\hat{Q}\hat{P}\right)
\nonumber \\
&=&\frac 12\left( \hat{P}+\frac{\lambda \omega }2\hat{Q}\right) ^2
+\frac 12\left( \omega ^2-\frac{\lambda^2\omega ^2}4\right) \hat{Q}^2.
\label{a19}\ea 
As $\lambda \omega $ is the friction coefficient, the new oscillational
frequency is just the correctly shifted one for a damped oscillator: 
\begin{equation}
\Omega =\sqrt{\left( \omega ^2-\frac{\lambda ^2\omega ^2}4\right) }.
\label{a20}
\end{equation}
Thus we find a consistent description of the ordinary Brownian motion
process in terms of continuous measurements.

\section{Conclusion} \label{SectConclud}

Non-minimally disturbing continuous quantum measurements were shown to
lead to dissipation of the measured system. The resulting model of
dissipation is characterized by a Lindblad master equation and thus
avoids the difficulties arising in other approaches. Although only
measuring a single observable has been considered in the present paper,
generalization on the case of measuring two or more observables is
straightforward.

\vskip 0.5cm
\centerline{\bf ACKNOWLEDGEMENT}\nopagebreak

One of the authors (M.B.M.) acknowledges support from the Royal
Swedish Academy of Sciences and the KTH 
where the main part of this work was completed. The work was supported
in part by the Deutsche Forschungsgemeinschaft and Russian Foundation
of Basic Research.

\end{document}